\documentclass[3p,times,twocolumn]{elsarticle}

\usepackage{amsmath}
\usepackage{url}

\usepackage{amssymb}
\usepackage{lineno}

\usepackage[figuresright]{rotating}

\begin{document}

\begin{frontmatter}

\title{Measurement of two­-particle pseudorapidity correlations in Pb+Pb collisions at $\sqrt{s_{NN}}$ = 2.76 TeV with the ATLAS detector}
\author{Sooraj Radhakrishnan on behalf of the ATLAS Collaboration}


\begin{abstract}
Two-particle pseudorapidity correlations, measured using charged particles with $p_{\mathrm{T}} >$ 0.5 GeV and $|\eta| <$ 2.4, from $\sqrt{s_{NN}}$ = 2.76 TeV Pb+Pb collisions collected in 2010 by the ATLAS experiment at the LHC are presented. The correlation function $C_N(\eta_1,\eta_2)$ is measured for different centrality intervals as a function of the pseudorapidity of the two particles, $\eta_1$ and $\eta_2$. The correlation function shows an enhancement along $\eta_- \equiv \eta_1 - \eta_2 \approx$ 0 and a suppression at large $\eta_-$ values. The correlation function also shows a quadratic dependence along the $\eta_+ \equiv \eta_1$ + $\eta_2$ direction. These structures are consistent with a strong forward-backward asymmetry of the particle multiplicity that fluctuates event to event. The correlation function is expanded in an orthonormal basis of Legendre polynomials, $T_n(\eta_1)T_m(\eta_2)$, and corresponding coefficients $a_{n,m}$ are measured. These coefficients are related to mean-square values of the Legendre coefficients, $a_n$, of the single particle longitudinal multiplicity fluctuations: $a_{n,m} = \langle a_na_m \rangle$. Significant values are observed for the  diagonal terms $\langle a_n^2 \rangle$ and mixed terms $\langle a_na_{n+2}\rangle$. Magnitude of $\langle a_{\mathrm{1}}^{\mathrm{2}} \rangle$ is the largest and the higher order terms decrease quickly with increase in $n$. The centrality dependence of the leading coefficient $\langle a_{\mathrm{1}}^{\mathrm{2}} \rangle$ is compared to that of the mean-square value of the asymmetry of the number of participating nucleons between the two colliding nuclei, and also to the $\langle a_{\mathrm{1}}^{\mathrm{2}} \rangle$ calculated from HIJING.
\end{abstract}


\end{frontmatter}

% \linenumbers

%\linenumbers
\section{Introduction}
\label{sec:intro}
Ultra relativistic heavy-ion collisions at RHIC and LHC create hot dense matter of deconfined quarks and gluons. The initial density distributions in the collisions fluctuate event to event and so a proper understanding of these fluctuations is necessary to describe and study the subsequent evolution of the produced matter. Study of the multiplicity correlations in the transverse direction and its fluctuations event by event have helped place important constraints on the transverse density fluctuations in the initial state \cite{phnx_vn:2011, atl_ebe:2013,atl_epc:2014}. Multiplicity correlations in the longitudinal direction are sensitive to the initial density fluctuations in pseudorapidty ($\eta$). These density fluctuations influence the early time entropy production and generate long-range correlations (LRC) which appear as correlations of the multiplicity of produced particles separated by large $\eta$ difference~\cite{bia_bzd_zal:2012, bzd_tea:2013, jia_huo:2014}. For example, EbyE differences between the number of nucleon participants in the target and the projectile may lead to a long-range asymmetry in the longitudinal multiplicity distribution \cite{bzd_tea:2013}. Longitudinal multiplicity correlations can also be generated during the final state, such as from resonance decays, jet fragmentation and Bose-Einstein correlations, which are typically localized over a smaller range of $\eta$ difference, and are commonly referred to as short-range correlations (SRC).

Previous studies of the longitudinal multiplicity correlations have focussed on the forward-backward correlations of the particle multiplicity in two symmetric $\eta$ windows around the center of mass of the collision system~\cite{bia_zal:2010, star_lrc:2009}. Recently, a more general method which uses on the correlation function in the full $\eta_1$, $\eta_2$ space has been proposed~\cite{bzd_tea:2013,jia_rad_zho:2015}. The orthogonal modes of the correlation function provides information on the orthogonal modes in EbyE single particle longitudinal multiplicity fluctuations and their magnitudes. The method had been applied to HIJING~\cite{gyu_wan:1994}  and AMPT~\cite{lin_ko_li_zha_pal:2005n} models to extract different shape components of the multiplicity fluctuation. In this proceedings, the correlation functions and the extracted magnitudes of the shape components of longitudinal multiplicity fluctuation, measured in Pb+Pb collisions at 2.76 TeV at the LHC using the ATLAS~\cite{atl_inst:2008} detector are presented.  

\section{Method and analysis procedure}
\label{sec:meth}
The two particle correlation function in pseudorapidity is defined as~\cite{vechvlad:2013}:
\begin{equation}
C(\eta_1,\eta_2) = \frac{\langle N(\eta_1)N(\eta_2) \rangle}{\langle N(\eta_1)\rangle\langle N(\eta_2)\rangle} 
\label{eq:meth1}
\end{equation}
where $N(\eta) \equiv \frac{dN}{d\eta}$ is the multiplicity density at $\eta$ in a single event and $\langle N(\eta) \rangle$ is the average multiplicity at $\eta$ for a given event class.  

In principle, the averages in Eq.~\ref{eq:meth1} should be calculated over event-classes defined using narrow centrality intervals, so that it contains only dynamical fluctuations that decouple from any residual centrality dependence of the average shape, $\langle N(\eta) \rangle$, which would lead to a modulation of the projections of the correlation function along the $\eta_1$ or $\eta_2$ axes. But due to experimental limitations and finite statistics, this modulation from the change of average shape cannot be completely removed in $C(\eta_1,\eta_2)$ as defined in Eq.\ref{eq:meth1}. However, these modulations can be removed by a simple redefinition of the correlation function~\cite{jia_rad_zho:2015}:
\begin{equation}
C_N(\eta_1,\eta_2) = \frac{C(\eta_1,\eta_2)}{C_P(\eta_1)C_P(\eta_2)} ,
\label{eq:meth2}
\end{equation}
where $C_P(\eta_{1(2)}) = \frac{\int C(\eta_1,\eta_2)d\eta_{2(1)}}{2Y}$, with $Y$ being the maximum value of $\eta_1$ and $\eta_2$. The resulting distribution is then renormalized such that the average value of $C_N(\eta_1,\eta_2)$ in the $\eta_1$ and $\eta_2$ plane is one. 

Following~\cite{bzd_tea:2013,jia_rad_zho:2015}, the correlation function is expanded into the orthonormal polynomials,
\begin{equation}
C_N(\eta_1,\eta_2) = 1+ \sum_{n,m=1}^{\infty} a_{n,m}\frac{T_n(\eta_1)T_m(\eta_2) + T_n(\eta_2)T_m(\eta_1)}{2},
\label{eq:meth3}
\end{equation}
where $T_n(\eta) = \sqrt{n+\frac{1}{2}}P_n(\eta/Y)$ and $P_n(x)$ are the Legendre polynomials. The $a_{n,m}$ coefficients in the expansion are related to the magnitudes of the shape fluctuations in the EbyE distribution, $a_{n,m} = \langle a_na_m \rangle$ where $a_n$ are the coefficients in the expansion of the single particle ratio, $\frac{N(\eta)}{\langle N(\eta) \rangle}  = 1 + \sum a_nT_n(\eta)$. 

In the analysis the correlation functions are calculated in 5\% centrality bins~\cite{atl_cent:2012}, defined using the transverse energy distribution in the ATLAS Forward Calorimeters (FCal). The analysis uses charged particle tracks reconstructed in the ATLAS inner detector to construct the correlation functions. The tracks are required to have $p_{\mathrm{T}} >$ 0.5 GeV and $|\eta| <$ 2.4. The correlation function is constructed as the ratio of distributions of same-event pairs ($S(\eta_1,\eta_2) \propto \langle N(\eta_1)N(\eta_2)\rangle$ and mixed-event pairs ($B(\eta_1,\eta_2) \propto \langle N(\eta_1)\rangle\langle N(\eta_2)\rangle$), $C(\eta_1,\eta_2) = \frac{S(\eta_1,\eta_2)}{B(\eta_1,\eta_2)}$. The events used for constructing the mixed event pairs are required to have similar total number of reconstructed tracks, $N_{\mathrm{ch}}^{\mathrm{rec}}$, (matched within 0.5\%) and $z$-coordinate of the collision vertex (matched within 2.5 mm). The events are also required to be close to each other in time to account for possible time-dependent variation of the detector conditions. To correct $S(\eta_1,\eta_2)$ and $B(\eta_1,\eta_2)$ for detector inefficiencies, the tracks are weighted by the inverse of their tracking efficiencies. Remaining detector effects largely cancel in the same to mixed event ratio. The systematic uncertainties in the correlation function are evaluated to be in the range 2$-$8.5\% depending on the centrality interval. These uncertainties are propagated into $C_N(\eta_1,\eta_2)$ and other derived quantities, as the $\langle a_na_m \rangle$. More details of the analysis and systematic uncertainties along with a complete set of results can be found in this reference~\cite{atl_lrc:2015}.

\section{Results}
\label{sec:res}
\begin{figure}
\includegraphics[width=1.0\columnwidth]{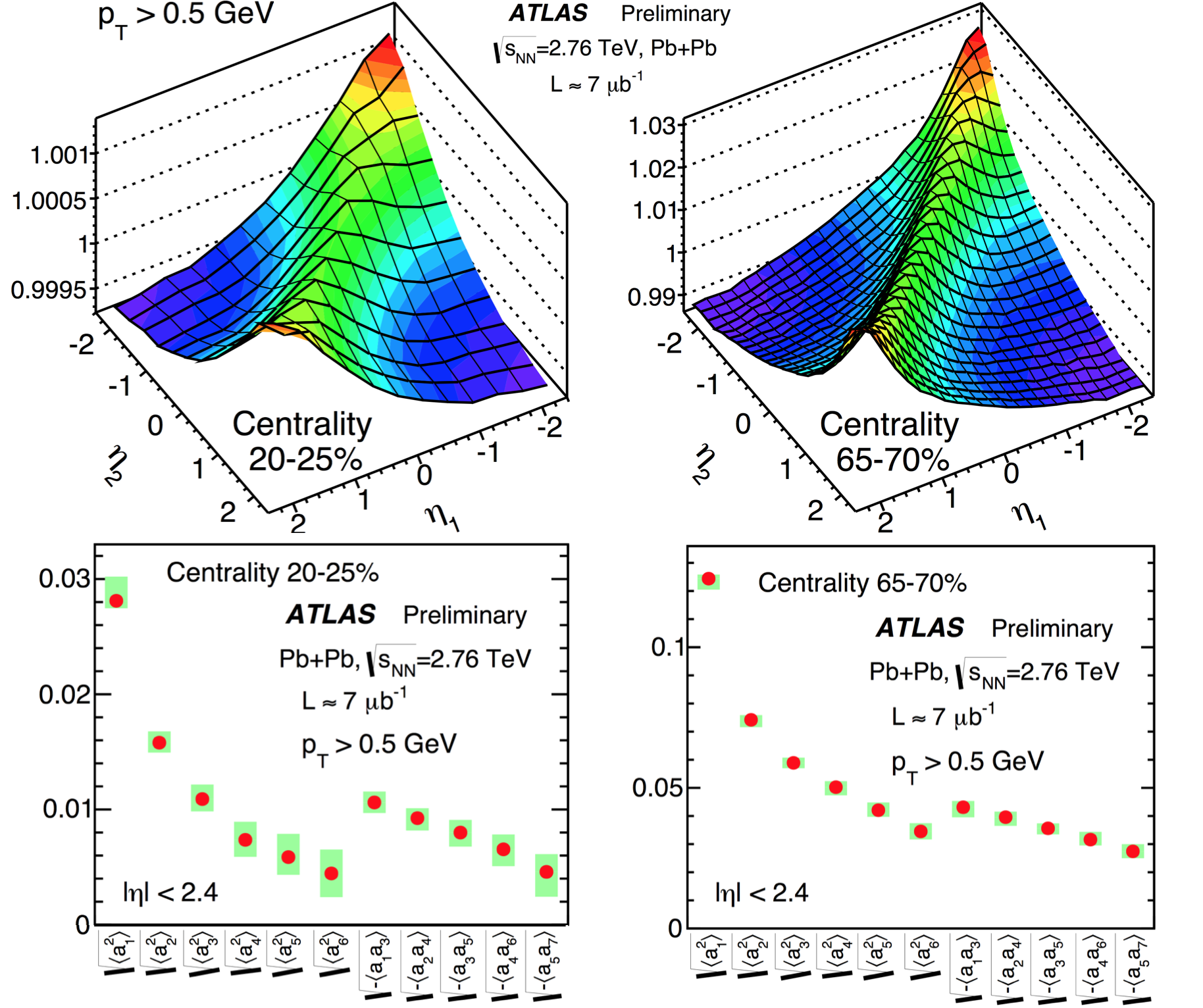}
\caption{Correlation function ($C_N(\eta_1,\eta_2)$) (top row) and the extracted $\langle a_na_m \rangle$ coefficients (bottom row) for 20-25\% and 60-65\% centrality classes ~\cite{atl_lrc:2015}.}
\label{fig:res1}
\end{figure}

The top panels of Figure.\ref{fig:res1} shows the correlation function $C_N(\eta_1,\eta_2)$ in a mid-central and a peripheral event class. The correlation functions show a broad ''ridge-like'' shape along $\eta_1 = \eta_2$, and a depletion in the large $\eta_-$ region around $\eta_1 = -−\eta_2 \approx \pm 2.4$. The magnitude of the ridge structure is larger in peripheral event classes than in central event classes and could have a significant contribution from SRC. The depletion in the large $\eta_-$ region reflects the contribution from the LRC. The lower panels of Figure.\ref{fig:res1} shows the extracted $\langle a_na_m \rangle$ coefficients for the two centrality classes. Non-zero values are observed for the diagonal terms, $\langle a_n^2 \rangle$ and also for mixed coefficients of the form $\langle a_na_{n+2} \rangle$. The first six diagonal terms and first five mixed terms are shown in the figure. Magnitude of $\langle a_{\mathrm{1}}^{\mathrm{2}} \rangle$ is much larger than the other coefficients and the magnitude of higher order terms decrease quickly with increasing $n$. The magnitudes of the coefficients are larger in the peripheral event classes. Also the magnitude of the higher order coefficients decrease less rapidly with increasing $n$ in the peripheral event class compared to the central event class. A significant contribution to $\langle a_na_m \rangle$, particularly the higher order terms, could be from short-range correlations which contribute to coefficients of all orders. 

To further analyse the features of the correlation function, the correlation function is expressed in terms of $\eta_- = \eta_1-\eta_2$ and $\eta_+=\eta_1+\eta_2$. The resulting correlation function, $C_N(\eta_{-−},\eta_+)$ is then projected onto the $\eta_{-−}$ ($\eta_+$) axis in narrow ranges of $\eta_+$ ($\eta_{−-}$). The shape of the projections along $\eta_-$ are more sensitive to the SRC as the SRC have a strong dependence on $\eta_-$. The shape of the projections along $\eta_+$ on the other hand is more sensitive to the long-range correlations. If the first-order coefficient $a_1$ is dominating, then the correlation function in $\eta_+$ and $\eta_-$ can be written as 
\begin{align}
%\begin{equation}
\label{eq:res1}
C_N(\eta_{-},\eta_+) \sim 1 + \langle a_{\mathrm{1}}^{\mathrm{2}} \rangle\frac{3}{8Y^2}(\eta_{+}^{2} - \eta_{-}^{2}) \\ \nonumber
\approx 1+0.065\langle a_{\mathrm{1}}^{\mathrm{2}} \rangle(\eta_+^2 - \eta_-^2)
\end{align}
%\end{equation} 
Since the SRC are expected to have a weak dependence on $\eta_+$, a quadratic dependence of the projection, $C_N(\eta_+)$, on $\eta_+$ can be expected. 

\begin{figure}
\includegraphics[width=1.0\columnwidth]{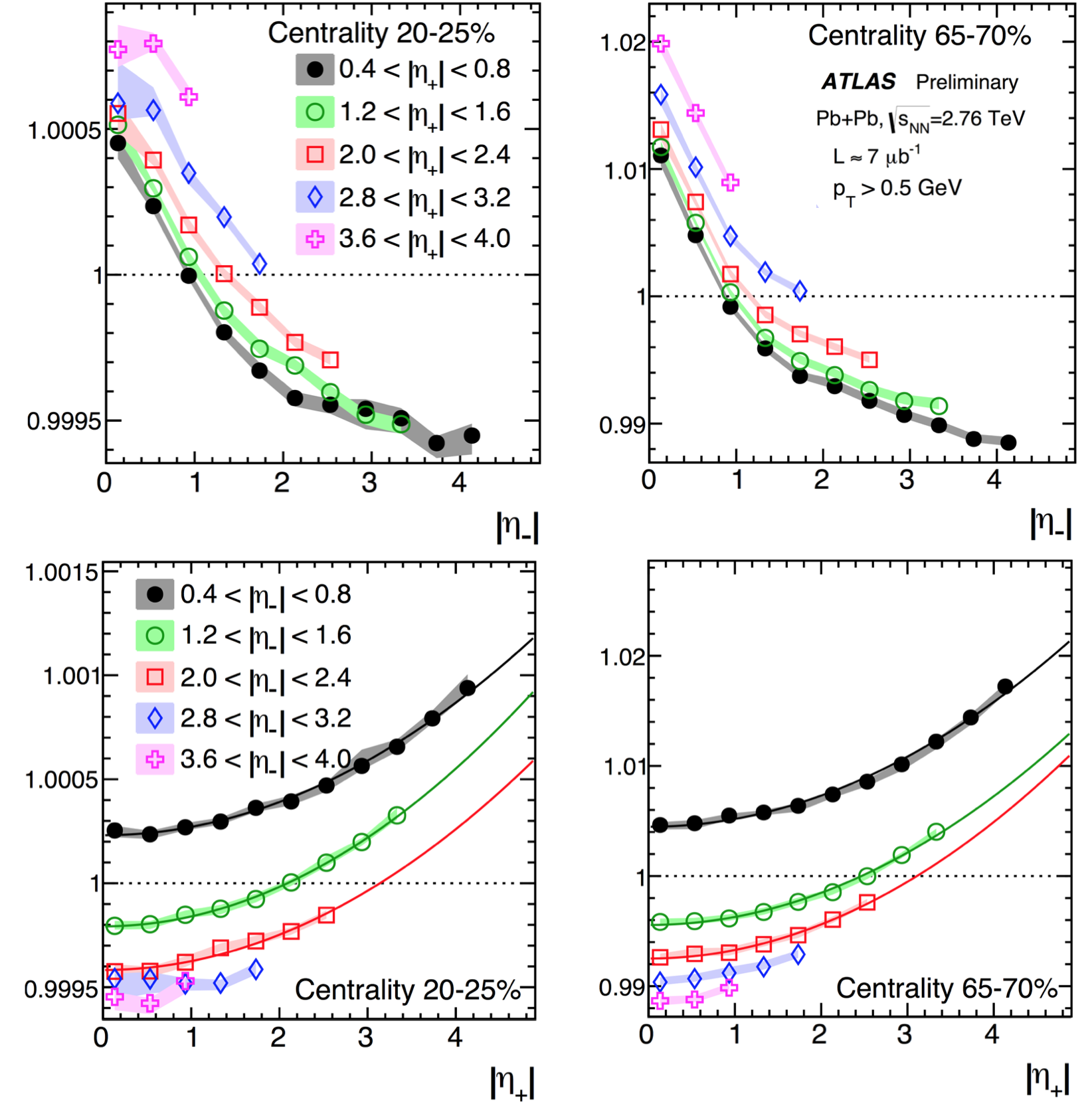}
\caption{$C_N(\eta_−)$ for different $\eta_+$ slices (top row) and $C_N(\eta_+)$ for different $\eta_-$ slices (bottom row) for 20-25\% and 60-65\% centrality classes~\cite{atl_lrc:2015}.}
\label{fig:res2}
\end{figure}

The projections, $C_N(\eta_{-−})$ and  $C_N(\eta_+)$, are shown in Figure.\ref{fig:res2} for different $\eta_+$ and $\eta_-$ slices respectively for the two centrality classes. Along the $\eta_-$ direction, the projections for all $\eta_+$ slices peak at $\eta_{−-}$ = 0 and decrease quickly for $\lvert \eta_{-}\rvert <$ 1, followed by a much weaker decrease beyond that. The quick decrease in $\lvert \eta_{-}\rvert< $ 1 is consistent with the dominance of short-range correlations, which are mostly centred around $\eta_- \approx$ 0. The weaker decrease at large $\lvert \eta_{-}\rvert$ could be related to the $-\eta_-^2$ term in Eq.~\ref{eq:res1}.

The projections along $\eta_+$ direction show a clear quadratic dependence on $\eta_+$ for all $\eta_-$ slices used for projection. This reflects the dominant $a_1$ component of the long-range correlation. The quadratic dependence is quantified by fitting the $C_N(\eta_+)$ data with the function $C_N(\eta_+) = 0.065\langle a_{\mathrm{1}}^{\mathrm{2}} \rangle\eta_+^2 + b$, where $b$ is a constant. The function fits the data quite well and are shown as solid lines in the figure. The $\langle a_{\mathrm{1}}^{\mathrm{2}} \rangle$ values are extracted from the fit for each $\eta_-$ window used for projection.

\begin{figure}
\includegraphics[width=1.0\columnwidth]{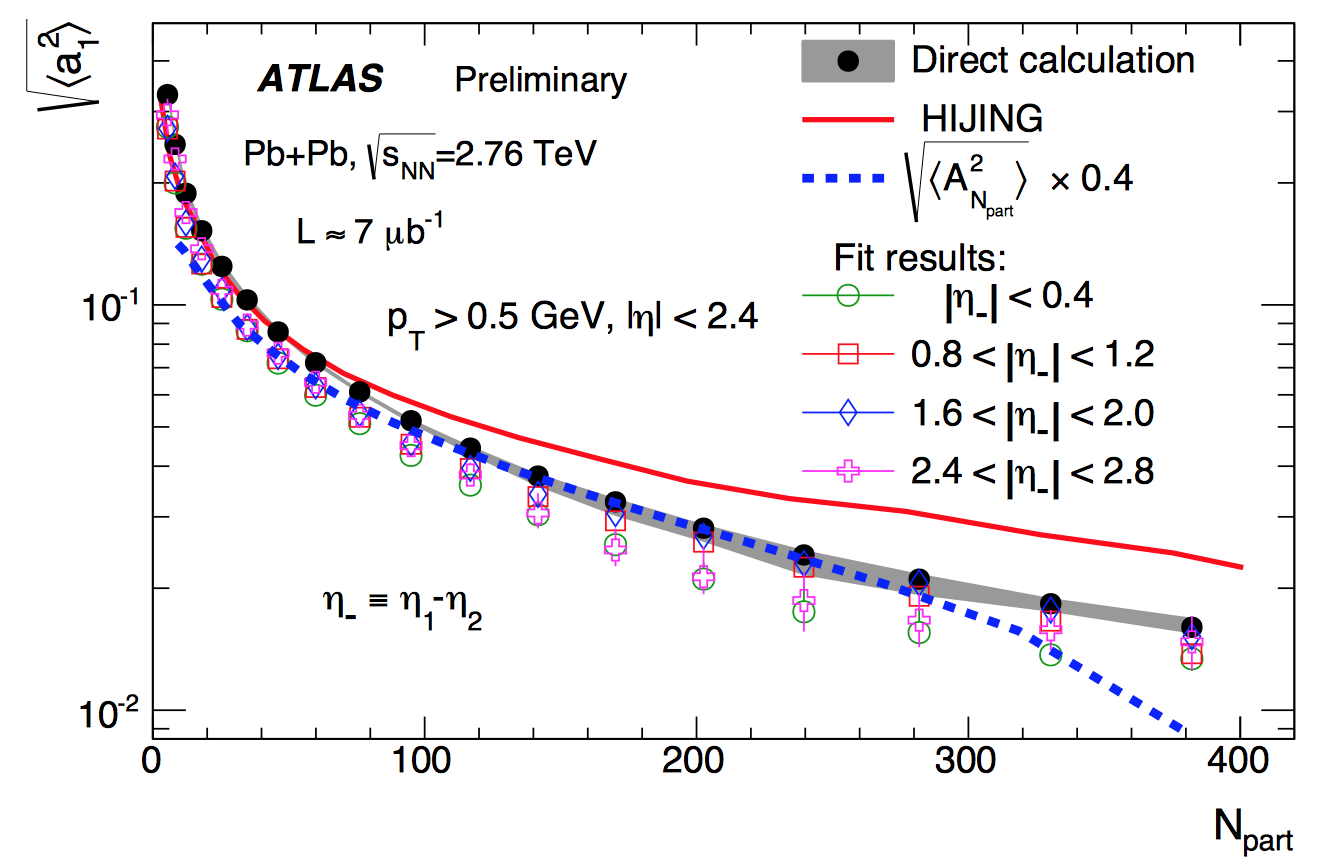}
\caption{$\langle a_{\mathrm{1}}^{\mathrm{2}} \rangle$ values calculated directly from the $C_N(\eta_1,\eta_2)$ (solid circles), obtained from fits shown in Fig.~\ref{fig:res2} (open symbols), HIJING model (solid line), as well as the RMS values of $A_{Npart}$ from Eq.~\ref{eq:res2} (dashed line). The shaded bands or error bars are the total uncertainties~\cite{atl_lrc:2015}.}
\label{fig:res3}
\end{figure}

Figure.\ref{fig:res3} shows the centrality dependence (in terms of the total number of participating nucleons, $N_{\mathrm{part}}$) of the RMS value of the first order coefficient, $a_1$. The values calculated from the Legendre expansion (Eq.~\ref{eq:meth3}) as well as those obtained from the fit to projections in Figure.\ref{fig:res2} are shown. The magnitude of $\langle a_{\mathrm{1}}^{\mathrm{2}} \rangle$ increase towards peripheral event classes. The values from the fits are always smaller than those from direct calculation by 2 $-–$ 20\% depending on the centrality interval and $\eta_{-}$ slice used for projection. This could be due to the contribution from the SRC to the values calculated using Legendre expansion.

Figure.\ref{fig:res3} also shows the centrality dependence of the RMS value of the asymmetry between number of forward going and backward going participants, $A_{N_{\mathrm{part}}}$ defined as
\begin{equation}
A_{N_{\mathrm{part}}} = \frac{N_{\mathrm{part}}^{\mathrm{F}} - N_{\mathrm{part}}^{\mathrm{B}}}{N_{\mathrm{part}}^{\mathrm{F}} + N_{\mathrm{part}}^{\mathrm{B}}},
\label{eq:res2}
\end{equation} 
where $N_{\mathrm{part}}^{\mathrm{F}}$ and $N_{\mathrm{part}}^{\mathrm{B}}$ denote the number of forward going and backward going participants respectively. The $A_{N_{\mathrm{part}}}$ values are calculated from a Monte-Carlo Glauber model~\cite{glaub:2007}. The RMS values, $\sqrt{\langle A_{N_{\mathrm{part}}}^2 \rangle}$, are scaled down by an arbitrary factor of 0.4 to approximately match the $\sqrt{\langle a_{\mathrm{1}}^{\mathrm{2}} \rangle}$ values. The centrality dependence of the RMS values of $A_{N_{\mathrm{part}}}$ quite well match the centrality dependence of the RMS values of $a_1$ in the mid-central classes, but show a stronger decrease in the most central event classes and a weaker increase in the more peripheral event classes. The good match between the centrality dependence of the RMS values of $a_1$ and $A_{N_{\mathrm{part}}}$ over a large centrality range suggest that $a_1$ modulations are driven by $A_{N_{\mathrm{part}}}$ and is consistent with the observation in~\cite{jia_rad_zho:2015} that EbyE, $a_1$ is strongly correlated with $A_{N_{\mathrm{part}}}$. Also shown in the figure are $\sqrt{\langle a_{\mathrm{1}}^2 \rangle}$ values calculated from HIJING. The values from HIJING over-estimate the data except in the most peripheral event classes. 

\section{Summary and conclusions}
Two-particle pseudorapidity correlation functions $C_N(\eta_1,\eta_2)$ are measured as a function of centrality for charged particles with $p_{\mathrm{T}} >$ 0.5 GeV and $|\eta| <$ 2.4, for Pb+Pb collisions at 2.76 TeV. The correlation function shows a ''ridge like'' enhancement along the $\eta_1 \approx \eta_2$, and suppression at $\eta_1 \approx −\eta_2 \sim \pm 2.4$. These structures are further investigated by projecting the 2-D correlation function as function of $\eta_-$ and $\eta_+$ in narrow ranges of $\eta_−$ and $\eta_+$ respectively. The $C_N(\eta_-)$ projections show strong contribution from the short-range correlations, particularly in the region $|\eta_-| <$ 1, where the magnitude of the correlation decrease quickly with increase in $|\eta_-|$. The $C_N(\eta_+)$ distribution shows a clear quadratic dependence, characteristic of a forward-backward asymmetry induced by the asymmetry in the number of participating nucleons in the two colliding nuclei.  

The correlation function is decomposed into a sum of products of Legendre polynomials those describe the different shape components, and the coefficients $\langle a_na_m \rangle$ are calculated. Significant values are observed for $\langle a_n^2 \rangle$ and $\langle a_na_{n+2} \rangle$. Magnitude of $\langle a_{\mathrm{1}}^{\mathrm{2}} \rangle$ is much larger than that of other coefficients and the magnitude of higher order terms decrease with increase in $n$. These coefficients are observed to increase for peripheral collisions, consistent with the increase of the multiplicity fluctuation for smaller collision systems. The centrality dependence of $\langle a_{\mathrm{1}}^{\mathrm{2}} \rangle$ is compared with the centrality dependence of the coefficient of the quadratic term in the fits to $C_N(\eta_+)$. The fit results are 2$–-$20\% smaller than that obtained from a Legendre expansion, which could be due to the stronger influence of short-range correlations on the values calculated directly from the Legendre expansion.

\section*{Acknowledgement}
\vspace{-0.1in}
This research is supported by NSF under grant number PHY--1305037
\vspace{-0.1in}



%% References

\nocite{*}
\bibliographystyle{elsarticle-num}
\bibliography{proc_hp2015}



\end{document}